# Highly-Secure Physically Unclonable Cryptographic Primitives Using Nonlinear Conductance and Analog State Tuning in Memristive Crossbar Arrays

Hussein Nili[1*], Gina C. Adam[1,3], Mirko Prezioso[1], Jeeson Kim[2], Farnood Merrikh-Bayat[1], Omid Kavehei[2†], and Dmitri B. Strukov[1]

[1] University of California Santa Barbara, Santa Barbara, CA 93106-9560, U.S.A.
[2] Royal Melbourne Institute of Technology University, Melbourne, Victoria 3000, Australia
[3] National Institute for R&D in Microtechnologies, Bucharest, Romania
Email: [*]hnili@ece.ucsb.edu, [†]omid.kavehei@rmit.edu.au, [§]strukov@ece.ucsb.edu

## Abstract

The rapidly expanding hardware-intrinsic security primitives are aimed at addressing significant security challenges of a massively interconnected world in the age of information technology. The main idea of such primitives is to employ instance-specific process-induced variations in electronic hardware as a source of cryptographic data. Among the emergent technologies, memristive devices provide unique opportunities for security applications due to the underlying stochasticity in their operation. Herein, we report a prototype of a robust, dense, and reconfigurable physical unclonable function primitives based on the three-dimensional passive metal-oxide memristive crossbar circuits, by making positive use of process-induced variations in the devices' nonlinear *I-V*s and their analog tuning. We first characterize security metrics for a basic building block of the security primitives based on a two layer stack with monolithically integrated 10×10 250-nm half-pitch memristive crossbar circuits. The experimental results show that the average uniformity and diffusivity, measured on a random sample of 6,000 64-bit responses, out of ~697,000 total, is close to ideal 50% with 5% standard deviation for both metrics. The uniqueness, which was evaluated on a smaller sample by readjusting conductances of crosspoint devices within the same crossbar, is also close to the ideal 50% ± 1%, while the smallest bit error rate, i.e. reciprocal of reliability, measured over 30-day window under ±20% power supply variations, was ~1.5% ± 1%. We then utilize multiple instances of the basic block to demonstrate physically unclonable functional primitive with 10-bit hidden challenge generation that encodes more than $10^{19}$ challenge response pairs and has comparable uniformity, diffusiveness, and bit error rate.

## Introduction

The advent of the information technology era has stimulated an unprecedented expansion of interconnected networks and devices. The sheer volume of personal and sensitive information continuously carried over shared and remotely accessible networks poses significant security challenges[1-3], which cannot be adequately addressed by conventional cryptographic approaches. Most conventional cryptographic approaches rely on "secret keys" stored in nonvolatile memories for data encryption and access authentication, which are vulnerable to physical and side-channeling attack including such direct probing and power analysis.[4,5]

As a result, security approaches based on physical hardware roots-of-trust have recently attracted significant attention. Somewhat analogous to biometric identifiers such as retinal and fingerprint imprints, hardware roots-of-trust are physically embedded with their cryptographic data through unique, individual structural properties that are virtually unpredictable and practically inimitable.[2,4,6-9] The cryptographic data should be immediately and reliably available upon interrogation, and effectively impossible to learn or extrapolate, even when challenged by aggressive model-building and machine learning attacks.[6,10] A class of hardware security primitives, Physical Unclonable Functions (PUFs), draw their cryptographic "keys" from fabrication process variations. Among the wide variety of proposed PUF architectures based on spatial variations in electronic hardware[3,11-19], resistive random access memory (RRAM) crossbar architectures are very promising due to their simple and relatively low-cost fabrication process, small footprint, low operational power, CMOS integration compatibility[20-25], and process-induced variations in *I-V* characteristics pertinent to the mixed electronic-ionic transport and ionic memory mechanism.[20-29]

The most accessible manifestation of process-induced compositional and structural spatial variations in RRAM arrays is the spatial variations of effective switching thresholds, i.e. the voltage at which device conductance is abruptly changed upon application of a ramping bias. A related example is spatial variations in the ON and OFF state conductances in the array upon application of a large voltage or current bias.[26-29] The physical source of these variations is arguably, the stochastic nature of filamentary conduction, arising from compositional inhomogeneity of the switching medium as well as variations in individual device profiles (e.g. electrode imperfections, random variations in surface roughness, etc.).[22,30-32] These "entropy" sources are hence, the almost exclusive foundation of the earlier proposed RRAM-based security primitives. [33-35] Majority of such proposals are theoretical and, moreover, the many proposed architectures assume a relatively large array size[9,26,28,35], extensive peripheral programming and control circuitry[28], and strict evaluation protocols (due to the relatively small signal-to-noise ratio of spatial state variations in optimized RRAM arrays) to achieve viable operational metrics. Furthermore, limiting the network operation to a binary memory array ignores the nonlinear transport *via* RRAM devices and reduces the array to a simple resistive network. The average distributions of high resistance and low resistance states likely follow a log-normal distribution and are typically similar for different spatial distribution and across different array instances.

Relying on such spatial entropy sources is therefore likely to pose challenges in creating unique security instances and largely ignores one of the main advantages of RRAM arrays – their analog programmability. Moreover, relying only on high/low resistance state distributions may result in a systematic bias towards one end of the side of the resistance distribution, thereby reducing the effective array size, regardless of control and evaluation schemes.

Spatial variations in binary states (i.e. nonuniformities in ON/OFF between array devices) are average, mesoscale exhibits of the inherent stochasticity of filamentary conduction mechanisms in RRAM devices. On the other hand, transient and non-linear characteristics of individual devices can potentially carry the embedded nanoscale stochasticity information with much higher fidelity.[36-40] This is specifically true for operation in higher resistance regimes, where current transport is dominated by nanoscale local compositional and structural defect structure and electronic defect structure.[41-47] The main caveats of this approach are the temporal stability and long-term reliability of network properties. While some of the hardware primitives may be tolerant to temporal fluctuations of device properties (such random function generators), majority of the security primitives, including PUFs, require a very high degree of stability for viable high throughput operations. We believe that unique opportunity is presented by nonlinear electron transport, which is typical for many valence-change (VCM) analog memristors. The nonlinearity depends on the programmed state and applied bias and is closely correlated to filamentary conduction's stochasticity.[23,48-51] It can therefore be utilized as a prominent source of the "entropy" in RRAM arrays, towards the development of robust security primitives.[52]

The core focus of this work is to utilize such inherent entropy of analog RRAM arrays to create robust and reconfigurable security primitives. We utilize monolithically integrated two-level stacks of CMOS compatible analog RRAM arrays with a large dynamic and finely tunable conductance range ($\sim$ 1 M$\Omega$ – 10 k$\Omega$) to design and experimentally verify the operation of nonlinear RRAM-based PUF primitives. We devise and verify an *ex situ* network tuning algorithm to optimize the "entropy space" of RRAM arrays. 384 kb-long strings of PUF response data are examined for randomness metrics, worst case reliability, and temperature stability. We further devise and experimentally evaluate instance reconfiguration strategies and introduce a highly robust challenge-obfuscated PUF network structure with improved metrics and information throughput.

**Tuning and operation of nonlinear PUF primitives**

Two-level stacks of monolithically integrated 10×10 analog RRAM arrays were employed for the design and implementation of nonlinear RRAM PUF network (Figure 1a). The two fully passive TiO$_{2-x}$ memristor crossbars with an active device area of $\sim$350 nm $\times$ 350 nm were fabricated using *in situ* low temperature reactive sputtering deposition, ion milling and a precise planarization step. The middle electrodes are shared between the bottom and top layers (Figure 1b, c). The fabrication flow ensures a high device yield (>95%) and low <175 °C temperature budget, compatible with CMOS integration. The devices in the two layers show large dynamic range (1 µS – 200 µS on average), strong nonlinearity (especially in lower conductance regimes) and low variations in high/low resistance states and effective switching thresholds (Figure 1d, e).. The devices in both layers can be continuously tuned within their dynamic range with $\sim$1% precision. The tuning results for 16 clearly distinguishable states equally spaced in the 2 µS – 32 µS regime illustrated in Figure 1f. Figure 1g demonstrates the bias-dependent nonlinearity when the devices are tuned in the 220±60 k$\Omega$ range. Fabrication and characterization details are further explained in our earlier report.[53]

To evaluate the performance of analog RRAM arrays as stand-alone security primitives, a single-layer PUF architecture is employed (Figure 2). The states of all devices in the RRAM array are first tuned to specific values (discussed below) to attain a spatially balanced random state distribution in the arrays. Single-bit binary response $b$ in the $M \times N$ crossbar circuit is calculated by biasing $m$ columns with voltage $V_B$ and collecting currents $I$ from $n$ virtually grounded rows and then comparing sum of all currents from one selected group of $n/2$ rows with sum of currents of the other (Figure 2a). The remaining (unselected) rows and columns in the RRAM array are kept either grounded or floating. A specific set of selected rows and columns, set of grounded unselected lines and set of unselected floating lines is determined uniquely by the input challenge. For such approach, the maximum number of challenge-response pairs in an array is $\binom{M}{m} \times \binom{N}{n} \times 2^{(M-m)+(N-n)}$, where the first term denotes the number of possible combinations for the selected devices and the second term refers to the number of floating and grounded configurations for unselected rows and columns. To strike a balance between response randomness and reliability in the proposed nonlinear PUF operation, we use a write-verify programming algorithm to keep device states within certain range[54-56] (Figure 2b), most optimal for the chosen PUF operation scheme and allowing for detectable current differential margin. Th algorithm assumes a simplified linear model of the analog RRAM network and serves as a computationally efficient weight determination scheme to eliminate inadvertent biases in spatial weight distribution across the network. As such, it can be used as the foundation for sophisticated network optimization algorithms based on more detailed models of nonlinear device and array characteristics.

The map of device conductances and the corresponding conductance distribution histogram are illustrated in Figure 2c, d. Figure 2e shows the average device conductances along all rows and columns of the crossbar, denoting a fairly uniform conductance distribution across the array. We use this distribution as the starting point in evaluating the PUF primitive functional performance.

**PUF operational metrics**

The most common metrics of randomness and reliability in security primitives are inter- and intra-instance Hamming weight and Hamming distance. In particular, uniformity ($\mathcal{UF}$) is defined as intra-response Hamming weight, while diffuseness ($\mathcal{DF}$) represents intra-PUF Hamming distance. These metrics assess the randomness of a single PUF instance. Bit-error-rate ($\mathcal{BER}$) represents inter-trial Hamming distance between responses to identical challenges and evaluates PUF's reliability. Another important metric is uniqueness ($\mathcal{UQ}$), which is the inter-PUF Hamming distance between responses to identical challenges and as such effectively a measure of PUF instance uniqueness.

To evaluate the PUF network performance, we use a selection scheme with $m = 5$ rows and $n = 2$ columns, while keeping the unselected lines floating, and measure the PUF response at 3 different bias points $V_B = 200$ mV, $400$ mV, and $600$ mV. The available number of challenge response pairs in this case according to (1) is 697,000. We generate 384 kb of response data at each bias point subject to exclusive random selections, to construct a statistically substantial set and order the results in 64-bit response packets (i.e. total of 6,000 64-bit packets for each bias point are generated). Figure 3a, b show $\mathcal{UF}$, $\mathcal{DF}$, and $\mathcal{BER}$ metrics of the network for different biases. While PUF randomness is already near-ideal for biases $V_B = 200$ mV, it improves further when using higher voltages, i.e. biasing device in strongly nonlinear regime. The reliability of the network also improves substantially at higher biases, which can be partially attributed to improved signal-to-noise ratio. More importantly, an evaluation of uniqueness between different bias responses reveals the unparalleled utility of random variations in conductance nonlinearity (herein referred to as nonlinearity entropy space) in analog RRAM arrays. Spatial distribution of nonlinearity varies significantly both as a function of individual device characteristics and operational bias (Figure 3c, d). This, in turn, leads to significant redistribution of selection and sneak path currents at different bias points. Thus, responses from the same network to the same input challenges exhibit a high $\mathcal{UQ}$ (Figure 3e). Most prominently, the average $\mathcal{UQ}$ between linear ($V_B = 200$ mV) and nonlinear ($V_B = 600$ mV) regimes is 44.8±6.9%, which indicates an almost completely reconfigured entropy space. Such property can be readily translated to more complex and robust PUF structures design, utilizing strategies such as bias scrambling and key reconfiguration, and increasing the response throughput by encoding nonlinearity differential information in the response.

Additionally, we evaluated the stability of network response at higher temperatures (Figure 3f). The sharp decline of $\mathcal{BER}$ at increased biases, from 16.36±3.1% at 200 mV to 5.93±2.59% at 600 mV, also indicates the benefits of nonlinear operation in RRAM-based PUF networks. The improved response stability can be explained by higher signal-to-noise ratio of the current differentials and more stable charge transport across the filamentary conduction paths.[22,49,57,58]

**PUF Reconfigurability**

A critical gauge of competence for hardware security primitives is instance uniqueness (UQ). Since process-induced compositional and structural mismatches in RRAM arrays, sufficient instance uniqueness is a general assumption in most RRAM-based PUF structures. Additionally, PUF instance reconfigurability through array weight re-distribution is considered as one of the major advantages of RRAM-based PUFs, where large RRAM arrays have a sufficiently large conductance variations entropy space to constitute unique instances upon reprogramming.[9,27,29,59] Nevertheless, no meaningful experimental verifications are so far reported.

We have experimentally evaluated the more stringent reconfigurability criteria in nonlinear analog RRAM PUFs through two strategies. In the first approach, we generate five different array weights distribution using the weight tuning algorithm to reprogram a single array (Figure 4a). We generate 32 kb responses for three different bias points on each distribution, subject to the same set of input challenges. The resulting UQ over these distributions have an ideal mean of 50% for all three bias points (Figure 4b). Moreover, response uniqueness distribution improves at higher biases, leading to a small variance of 0.9% at 600 mV. This indicates the higher sensitivity of the nonlinear response to spatial variation and increased robustness of instance randomness in nonlinear operations.

In the second strategy, we generate 10 different distributions using a device "rattling" routine with considerably lower programming time and power. In this approach, we apply short (10 μS) RESET pulses with randomly assigned amplitudes and a maximum amplitude of 70% average RESET threshold, to each device. The resulting conductance distributions are illustrated in Figure 4c. For these distributions, the UQ significantly improves at nonlinear operational points. While the inter-PUF UQ for linear 200 mV operation is 24.8 ±6.3%, it is upgraded to the near ideal point of 50.07±2.1% for nonlinear 600 mV operation. These results further highlight the utility of embedded RRAM nonlinearity in hardware security primitives. In this case, analog RRAM array can be reconfigured through a fast and low power step to a completely unique instance, operated at its nonlinear regime. Nonetheless, the high bias and spatial sensitivity of RRAM nonlinearity distribution can be utilized to devise even more robust and power efficient reconfiguration methods.

**NL-RPUF structure: a robust high throughput security primitive**

Finally, we introduce and experimentally verify the operation of an easily reconfigurable challenge-response obfuscated architecture, that takes full advantage of the inherent nonlinearity and spatial distribution entropies in analog memristive arrays. The Nonlinear Resistive PUF (NL-RPUF) separates a large crossbar array or an array collective, into multiple challenge input layer ($CBA_i$) and response output Layer ($CBA_o$) segment arrays (Figure 5a). The two layers contain operating as well as dummy segment arrays which do not contribute to the PUF response and only scramble the network's power profile. The challenge is applied to segment arrays in the input layer to construct an $l$-bit hidden challenge (HC), which is then applied to segment arrays in the output layer to generate an $n$-bit response. For an optimum network operation, each segment array must exhibit excellent stand-alone randomness and reliability metrics. Considering each input segment array is independently selected and operated for the generation of a single hidden bit, and assuming the same number of row and columns are selected on each segment array, the nominal input space will be $\left( \binom{M_{D_i}}{m_i} \times \binom{N_{D_i}}{n_i} \right)^l$, where $M_{D_i}$, $N_{D_i}$, $m_i$ and $n_i$ are the total and selected rows and columns, respectively. The obfuscation of challenge-response correlation strengthens the PUF structure against direct probe and side-channeling attacks. Moreover, the transfer function for the proposed for the NL-RPUF is a composition of $CBA_i$ and $CBA_o$ transfer functions, $f_{NL-RPUF} = f_{CBA_o}(f_{CBA_i})$, which adds to the complexity of the transfer function and thus, likely improves the network resilience against various modeling attacks.

To demonstrate the basic performance of the NL-RPUF structure, we have implemented a simplified network with ten input and one output segment arrays. Each segment array in $CBA_i$ generates 1 hidden bit to construct a 10-bit HC that is used to address the rows and columns in $CBA_o$ and generate a 1-bit response. All segment arrays are implemented sequentially on the same 3D analog array using distributions generated by the "rattling" redistribution strategy (Figure 4c). Challenges to the input layers are the kept the same for generation of each HC to track the NL-RPUF response uniqueness in a worst-case selection scenario. 32 kb response bits to exclusive random inputs are generated at three bias point (200-600 mV) and ordered in 64-bit packets to construct 500 response keys. The uniformity and reliability of the response keys show marked improvements over the single layer network and the inter-bias response uniqueness significantly improves in the NL-RPUF network, yielding largely unique responses at different bias levels from the same network. The nonlinear network allows for multiple strategies to increase the input space and response throughput of the PUF instance. One such approach is employing the bias level tuning as a $k$-bit input to increase the challenge-response pair space. This is contingent on the high RRAM devices' nonlinearity and inter-bias uniqueness in NL-RPUF and will increase the challenge-response pair space by a factor of $2^k$. Here we have employed a 3-bit input to tune the bias point in 200-600 mV range in both layers simultaneously, and generate 500 multi-bias 64-bit response keys. The randomness and reliability metrics for the bias-encoded keys remain completely robust (Figure 5b) while exhibiting an average 26.8% uniqueness from single bias responses. The resulting scrambling of the current read-out profile will also complicate power-monitoring and probing attacks.

The high functional nonlinearity can also be utilized to upgrade the response bit level in NL-RPUF. We have combined the linear and nonlinear responses (at 200 and 600 mV, respectively) to generate quaternary (or 2-bit binary) responses (Figure 5c). In this case the nominal output space is increased by a power of two, almost completely preventing brute-force and model-building attacks. The excellent randomness and reliability characteristics of quaternary bits further confirms the unparalleled utility of nonlinearity entropy in constructing robust and secure physical functions.

## Conclusions

We have detailed the design and successful implementation of highly robust and reconfigurable NL-RPUF security primitives based on nonlinear analog RRAM arrays, utilizing two-level stacks of monolithically integrated 10×10 analog RRAM arrays. Utilizing stable intrinsic entropy space most directly correlated with nanoscale filamentary conduction processes in VCM analog memristors, namely, high-precision conductance tuning and state/bias-dependent conductance nonlinearity, we have demonstrated NL-RPUF's excellent operational metrics, its unique customizability between high security and power efficiency, and unparalleled fast and cost/power-efficient instance reconfigurability. We have further showcased the design and operation of a double-layer NL-RPUF structure that notionally offers high resilience against invasive and non-invasive physical and model-building attacks, as well as excellent security metrics and the capacity to massively increase the primitive's information throughput. These results indicate that such profound controllable and re-programmable nonlinearity features in analog VCM devices are an impressively ubiquitous asset in design and operation of next generation hardware security primitives. To the best of our knowledge, no comparable property has been reported in conventional mainstream CMOS computing devices or other types of emerging nonvolatile memory devices.

H. Nili *et al.*, "Highly-Secure PUFs Based on Memristive Xbar Arrays", Nov. 2016.

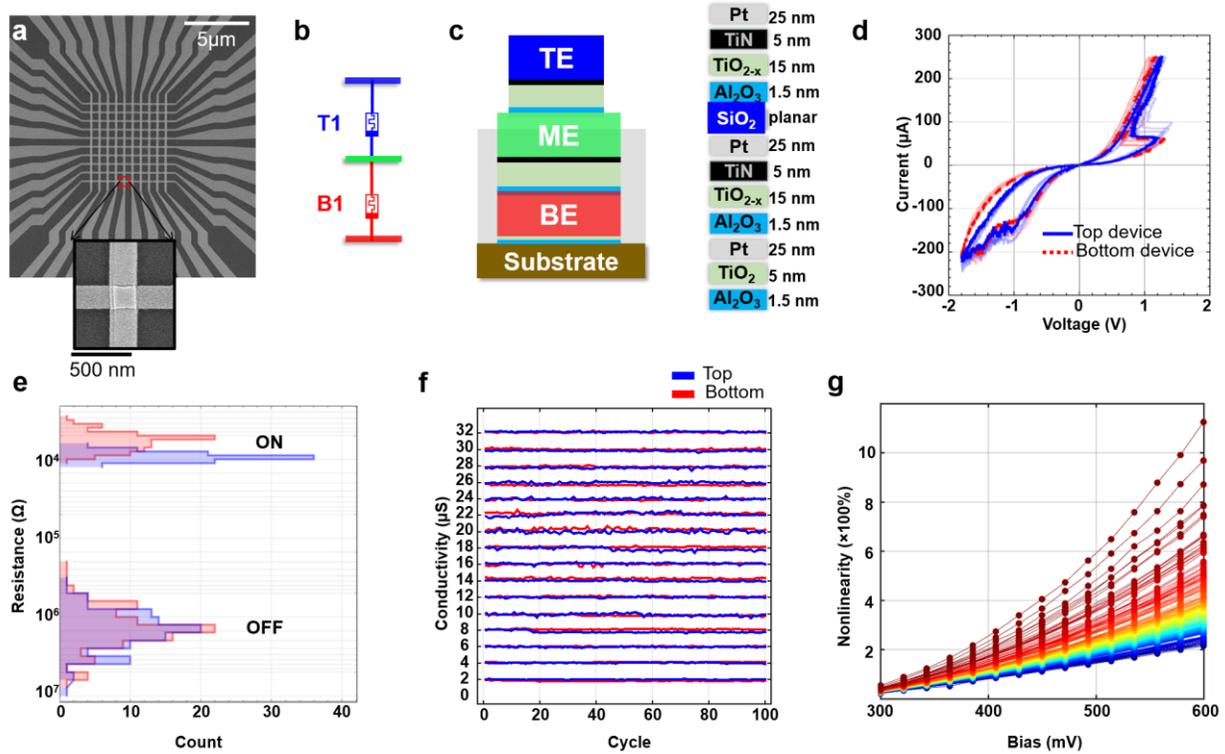

**Figure. 1**. Structural and operational characteristics of two-layer monolithically stacked 3D analog RRAM crossbar arrays: (a) Top-view SEM image of the 3D stacked crossbar. (b) Equivalent circuit and (c) cross-sectional schematic of the stack denoting material layers and thicknesses. (d) *I-V* curves for all 2×10×10 devices with two representative curves being highlighted for comparison. (e) Cumulative histogram for the devices' ON and OFF state conductances, measured at 0.3 V. (f) Tuning results to 16 different conductive states for top and bottom devices, equally spaced from 2 μS to 32 μS. (g) Nonlinearity at $V_0 = 200$ mV, calculated as a ratio of |1-$V/V_0×I(V_0)/I(V)$|/100, for all 2×10×10 devices tuned to 220±60 KΩ at 200 mV.

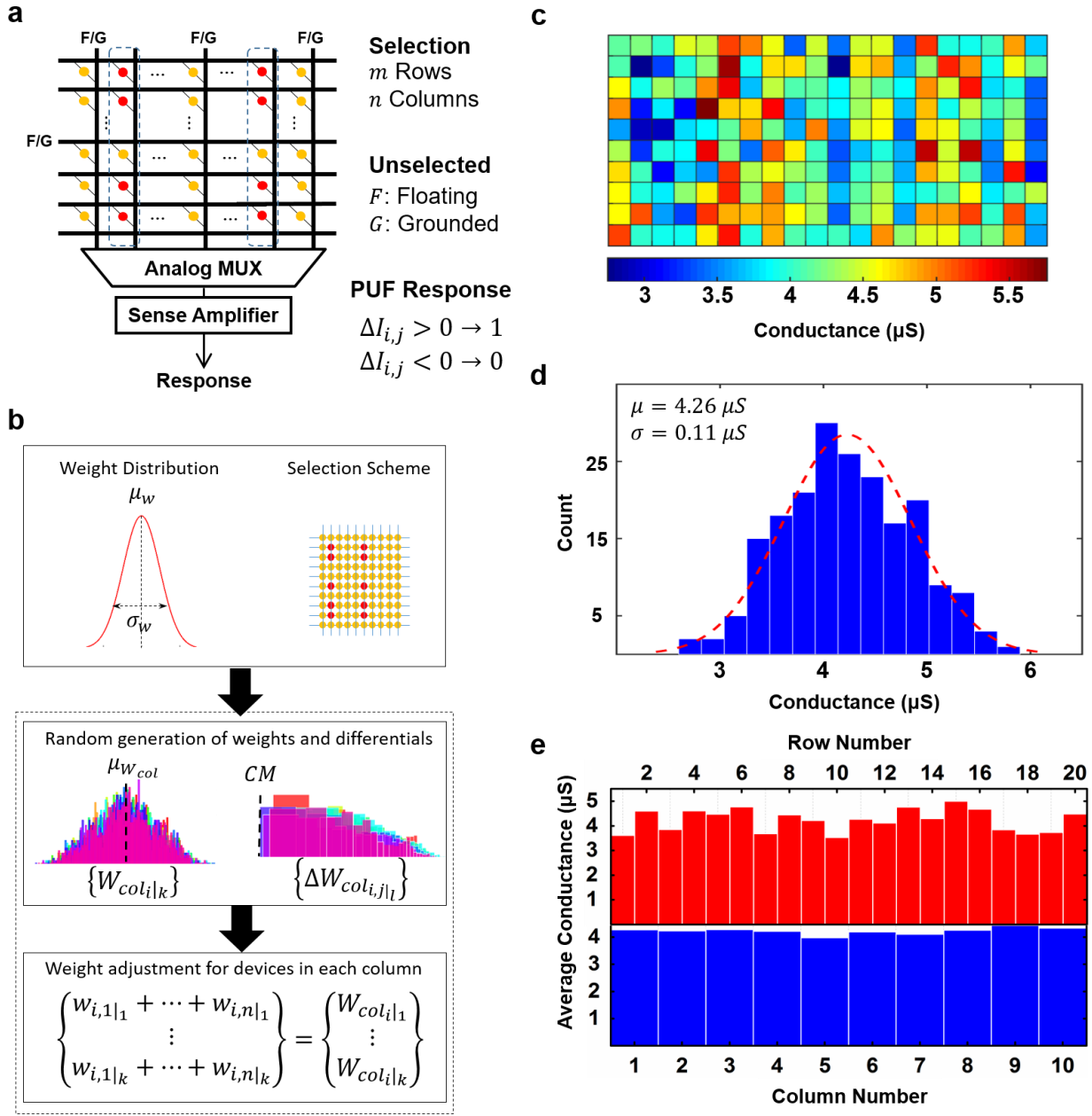

**Figure 2.** (a) Single-layer analog RRAM crossbar PUF architecture. The current differential between the selected $m$ rows and $n$ columns (with the rest of the network either floating or grounded) generates a 1-bit PUF response. (b) Distribution generation algorithm: Target (desired) array weight distribution ($\mu_w$, $\sigma_w$) and selection scheme are fed to a random column weight and differential generator to attain a statistically significant set (satisfying mean weight and confidence

margin criteria) which is then used to solve $k$ linear equations to adjust weights for devices on each column. (c) Spatial distribution, (d) average column and row conductances, and (e) histogram of conductance distribution at 200 mV of a tuned 3D stacked double array.

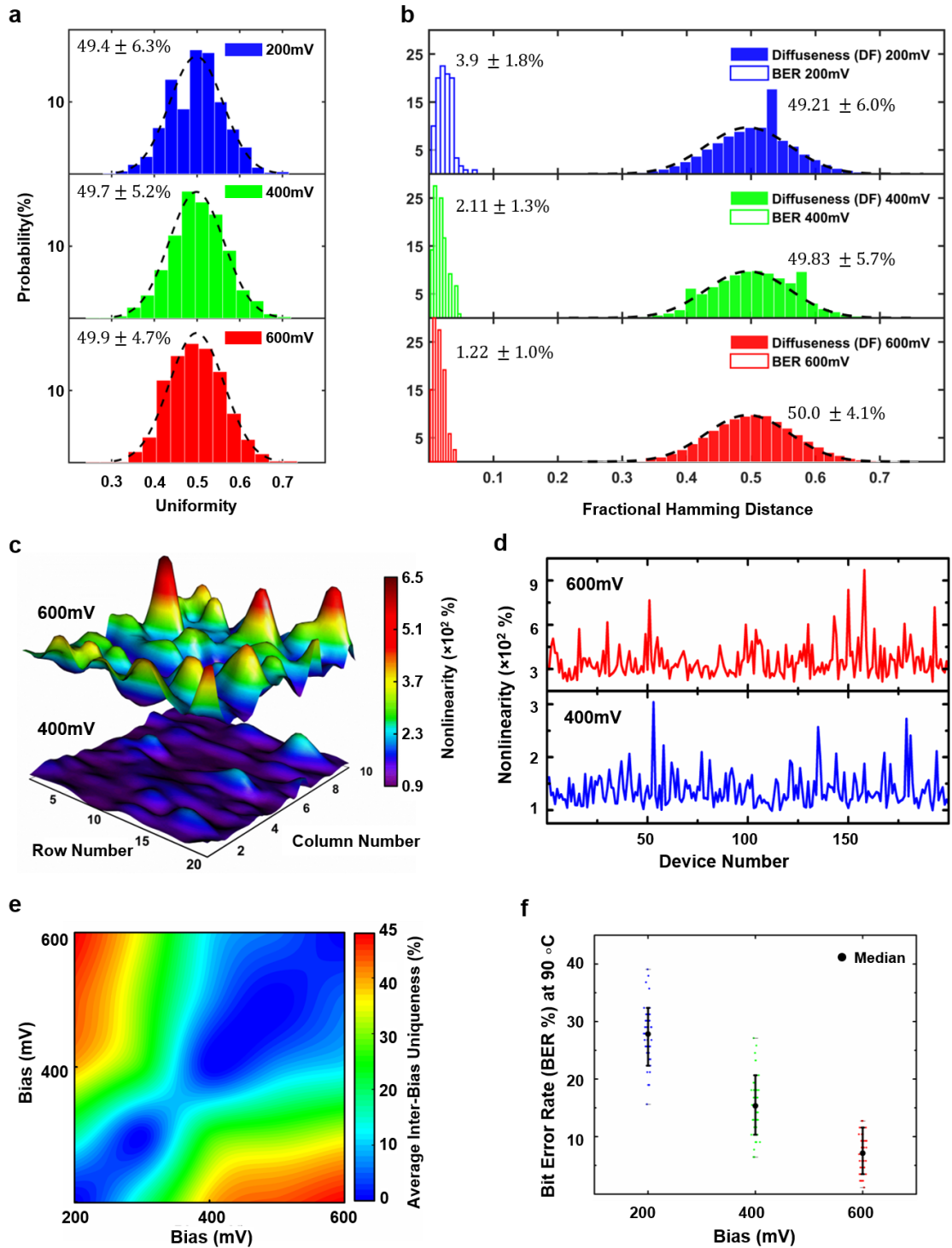

**Figure 3.** Operational metrics of the proposed PUF circuit based on three-dimensional analog RRAM arrays. (a) PUF response keys uniformity at different biases. (b) Response keys diffuseness and bit error rate at different biases. The bit-error rates are calculated by monitoring 16 kb random

challenge response pairs over a 30-day window in 10-day intervals, subject to constant bias stress and random supply source variations (±20%) to account for ageing and environmental factors. (c) Spatial distribution of nonlinearity at 400 mV and 600 mV biases. (d) Data from panel c shown as a line plot. (e) Contour map of inter-bias uniqueness between the responses generated for the same challenges at different biases. (f) Bit error rate for 4800 response bits at 90 °C at three different biases. The temperature is slowly ramped up to the target value and is kept constant for a period of 30 minutes before measurement start, for a total time length of 3 hours.

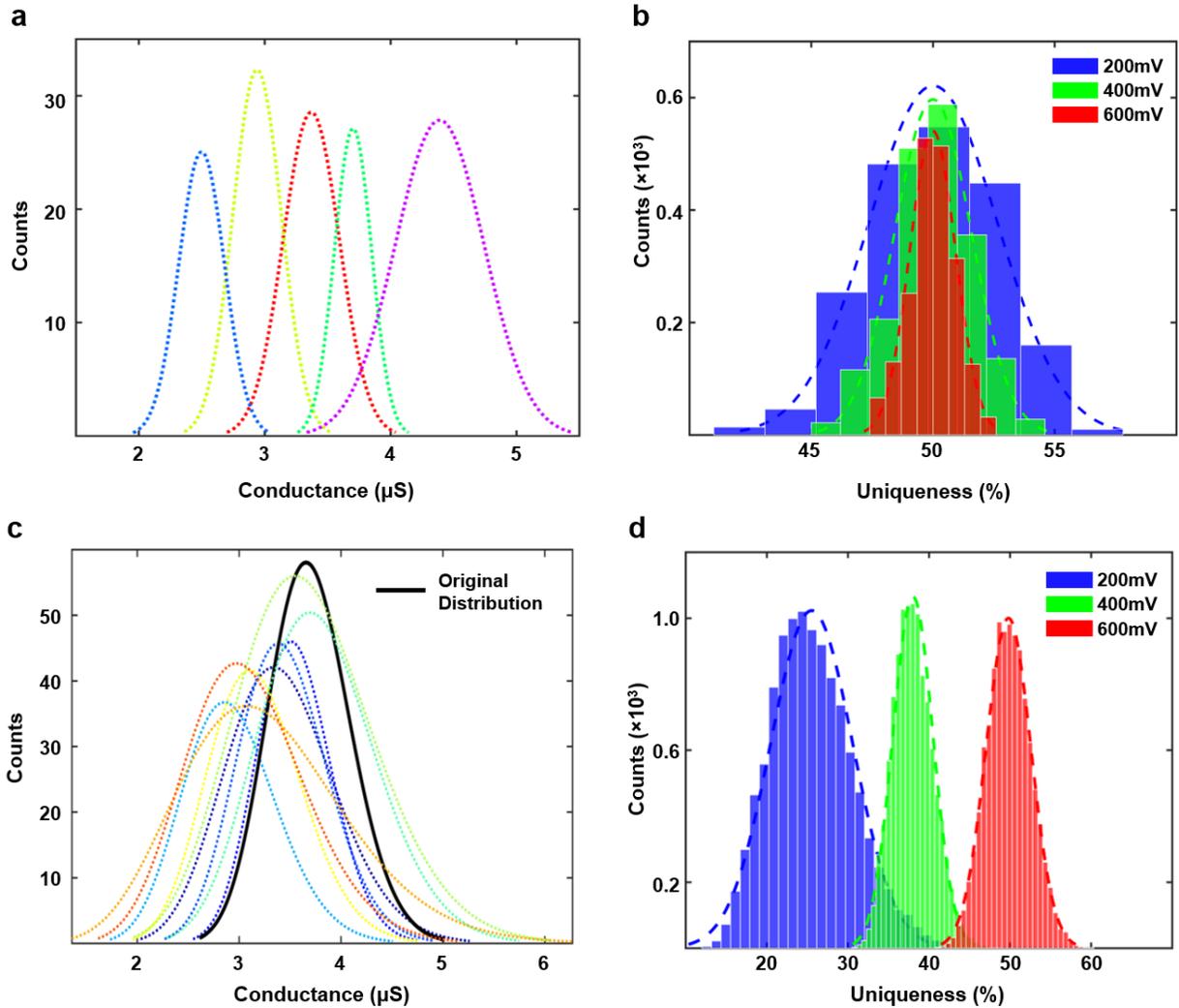

**Figure 4.** (a) Five independent analog RRAM distributions implemented on the same array using the tuning algorithm. (b) Inter-PUF UQ for array distributions in (a). The uniqueness distributions are 50.01±2.65 at 200 mV, 50.03±1.75 at 400 mV, and 49.99±0.9 at 600 mV operations (c) Ten array distributions using the rattling redistribution scheme. (d) Inter-PUF UQ for rattling strategy. (b). The uniqueness distributions are 24.8±6.3 at 200 mV, 38.2±3.3 at 400 mV, and 50.07±2.1 at 600 mV operation.

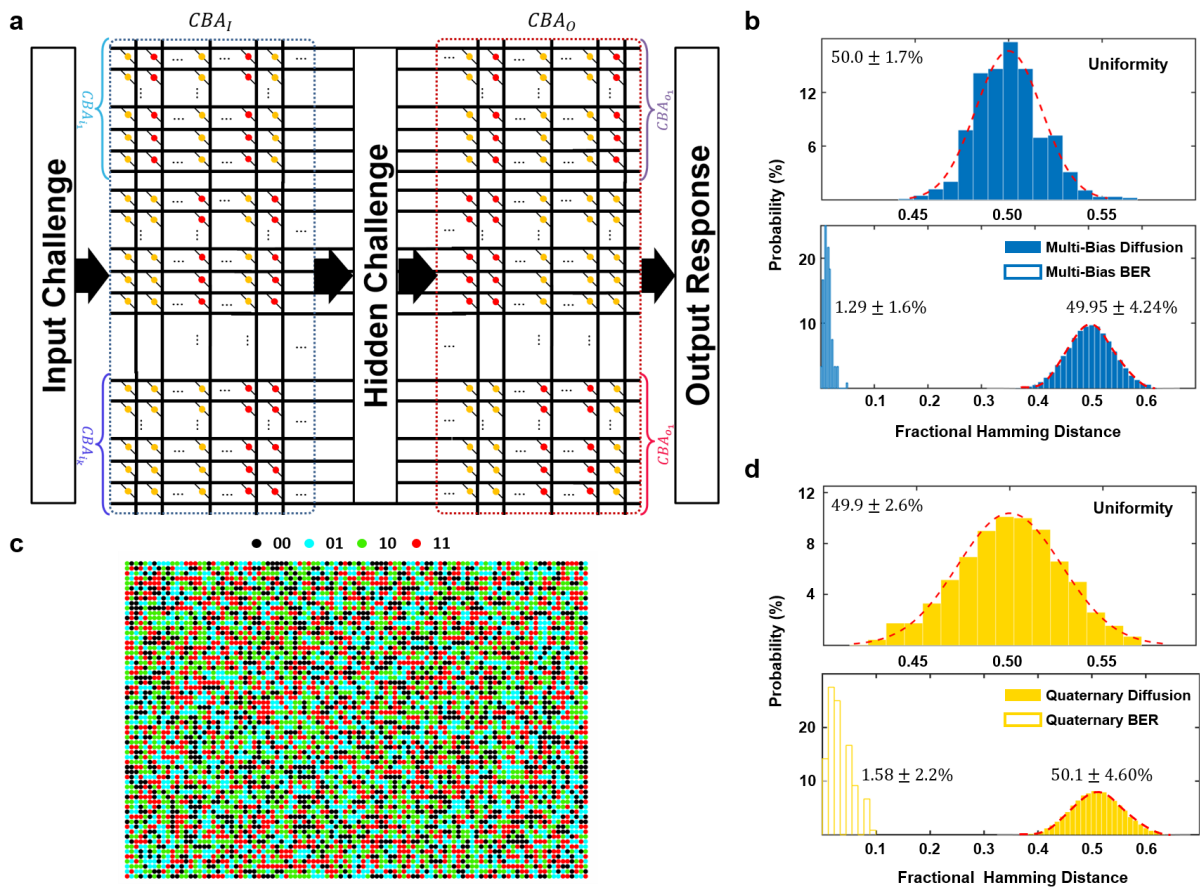

**Figure 5.** (a) Schematic of the operation of NL-PUF architecture. (b) Inter- and intra-response randomness metrics for multi-bias response bits. (c) A map of hundred 64-bit quaternary NLRPUF response keys. (d) Inter- and intra-response randomness metrics for quaternary response bits.

H. Nili *et al.*, "Highly-Secure PUFs Based on Memristive Xbar Arrays", Nov. 2016.